\documentclass[11pt]{article}
\usepackage{amssymb}
\usepackage{amsmath}
\usepackage{amscd}
\usepackage{latexsym}

\catcode`@=11
\renewcommand{\section}{\@startsection{section}{1}{0pt}{\medskipamount}
{\medskipamount}{\large\bf}}
\numberwithin{equation}{section}
\catcode`@=12
\def\a{\alpha}

\def\D{\Delta}
\def\de{\delta}

\def\ve{\varepsilon}
\def\h{\eta}
\def\th{\theta}

\def\la{\lambda}
\def\m{\mu}
\def\n{\nu}
\def\r{\rho}
\def\s{\sigma}

\def\j{\psi}

\def\La{\Lambda}

\newcommand{\C}{\mathbb C}
\newcommand{\R}{\mathbb R}
\newcommand{\Z}{\mathbb Z}
\newcommand{\Hbb}{{\mathbb H}}
\newcommand{\Hcal}{{\cal H}}
\newcommand{\Pcal}{{\cal P}}

\def\ic{\mbox{i}}
\def\N2{$N{=}2$}
\def\pa{\mbox{$\partial$}}
\def\diff{\mbox{d}}
\def\tr{{\rm tr}}
\def\sfrac#1#2{{\textstyle\frac{#1}{#2}}}
\def\>{\rangle}
\def\<{\langle}
\def\+{\dagger}
\def\={\ =\ }

\begin{document}

\begin{titlepage}
\setcounter{page}{0}
\begin{flushright}
hep-th/0502117\\
ITP--UH--02/05\\
\end{flushright}

\vskip 2.0cm

\begin{center}

{\Large\bf Noncommutative Instantons in 4k Dimensions}

\vspace{15mm}

{\Large
Tatiana A. Ivanova${}^\+$ \ and \ 
Olaf Lechtenfeld${}^*$ }
\\[10mm]
\noindent ${}^\+${\em Bogoliubov Laboratory of Theoretical Physics, JINR\\
141980 Dubna, Moscow Region, Russia}\\
{Email: ita@thsun1.jinr.ru}
\\[10mm]
\noindent ${}^*${\em Institut f\"ur Theoretische Physik,
Universit\"at Hannover \\
Appelstra\ss{}e 2, 30167 Hannover, Germany }\\
{Email: lechtenf@itp.uni-hannover.de}

\vspace{20mm}

\begin{abstract}
\noindent
We consider Ward's generalized self-duality equations 
for U($2r$) Yang-Mills theory on $\R^{4k}$ and their 
Moyal deformation under self-dual noncommutativity. 
Employing an extended ADHM construction we find two kinds 
of explicit solutions, which generalize the 't~Hooft and 
BPST instantons from $\R^4$ to noncommutative $\R^{4k}$. 
The BPST-type configurations appear to be new even 
in the commutative case.

\end{abstract}

\end{center}
\end{titlepage}

\section{Introduction and summary}

\noindent
Moyal-type deformations of gauge field theories arise naturally
in string theory when D-branes and an NS-NS two-form background are present
\cite{Sei}. Such noncommutative extensions are also interesting by themselves 
as specific nonlocal generalizations of ordinary Yang-Mills theories.
In particular, the question whether and how the known BPS classical solutions
(instantons, monopoles, vortices)~\cite{BPST}--\cite{Taubes} can be deformed
in this manner has been answered in the affirmative~\cite{NS}--\cite{Ham}.

In space-time dimensions higher that four, BPS configurations can still be
found as solutions to first-order equations, known as generalized self-duality
or generalized self-dual Yang-Mills (SDYM) equations. 
Already more than 20 years ago such equations were proposed~\cite{CDFN,W}, 
and some of their solutions were found e.g.~in~\cite{W}--\cite{P}. 
Also the ADHM solution technique~\cite{ADHM} was generalized from 
four-dimensional to $4k$-dimensional Yang-Mills (YM) theory~\cite{CGK}.
More recently, various BPS solutions to the noncommutative Yang-Mills equations
in higher dimensions and their brane interpretations have been investigated
e.g.~in~\cite{Wi}--\cite{PSW}. Some of these works employ the (generalized) 
ADHM construction extended to the noncommutative realm.
    
In this letter we use the extended ADHM method to construct new generalizations 
of the 't~Hooft as well as of the BPST instanton for U($2r$) gauge groups 
on self-dual noncommutative $\R^{4k}$. Only for simplicity we restrict 
ourselves to the gauge groups U(2) on $\R_\th^{4n+4}$ and U(4) on $\R_\th^{8}$,
respectively. In the first ('t~Hooft-type) case, our solutions do not have a 
finite topological charge, but their four-dimensional ``slices'' coincide with
the noncommutative $n$-instanton configurations of~\cite{ILM} featuring
noncommutative translational moduli. In the second (BPST-type) case, we correct
a faulty ansatz of~\cite{CGK} and find solutions to the (self-dually) deformed
generalized self-duality equations. Even in the commutative limit these 
configurations are novel instantons in eight dimensions.

\bigskip

\section{Quaternionic structure in $\R^{4k}$}

\noindent
We consider the $4k$-dimensional space $\R^{4k}$ with the metric 
$\de_{\mu\nu}$, where $\mu,\nu,\ldots=1,\ldots,4k$. 
Its complexification $\C^{4k}$ splits
into a tensor product, $\C^{4k}\cong \C^{2k}\otimes\C^{2}$. 
Complex coordinates $x^\mu$ on $\C^{4k}$ are related to complex
coordinates $x^{aP}$ on $\C^{2k}\otimes\C^{2}$ by~\cite{W, CGK}
\begin{equation}\label{x}
x^{\mu}\=V^{\mu}_{aP}\,x^{aP} \qquad\textrm{for}\quad
a=1,\ldots,2k \quad\textrm{and}\quad P=1,2\ ,
\end{equation}
where $(V^{\mu}_{aP})$ is a nonsingular invertible matrix which defines
a mapping from $\C^{2k}\otimes\C^{2}$ to $\C^{4k}$.
Imposing the reality condition $\overline{x^\mu} =x^\mu$, we obtain coordinates
on $\R^{4k}$. To preserve (\ref{x}) one should impose the following reality
conditions on $V^{\mu}_{aP}$ and $x^{aP}$:
\begin{equation}\label{V}
\overline{V^{\mu}_{aP}}\=\ve^{ab}\ve^{PQ}V^{\mu}_{bQ}\qquad\textrm{and}\qquad
\overline{x^{aP}}\=\ve_{ab}\ve^{}_{PQ}x^{bQ}
\ ,
\end{equation}
where
\begin{equation}\label{e}
\ve \= (\ve^{}_{PQ})=\begin{pmatrix}
\phantom{-}0&1 \\ -1 & 0\end{pmatrix} \qquad\textrm{and}\qquad
(\ve_{ab})\=\begin{pmatrix}\ve& &{\bf 0}_2 \\ &\ddots&\\ {\bf 0}_2& &\ve
\end{pmatrix}
\end{equation}
while the inverse matrices are defined by 
\begin{equation}\label{ee}
\ve^{PQ}\ve^{}_{QR}\=\de^P_R\qquad\textrm{and}\qquad\ve^{ab}\ve_{bc}\=\de^a_c
\ .
\end{equation}

The space $\R^{4k}$ with metric
\begin{equation}\label{ds}
\diff s^2\=\de_{\mu\nu}\,\diff x^\mu\diff x^\nu
\=\ve_{ab}\,\ve^{}_{PQ}\,\diff x^{aP} \diff x^{bQ}
\end{equation}
can be decomposed into a direct sum of $k$ four-dimensional spaces,
\begin{equation}\label{R4k}
\R^{4k}\ \cong\ \R^{k}\otimes\R^{4}\=\R^{k}\otimes\Hbb \=\Hbb^k\ ,
\end{equation}
with coordinates $x^{\mu_{\hat\imath}}$, where $\mu_{\hat\imath}=
(4\hat\imath{+}1,\ldots,4\hat\imath{+}4)$ is an index-quadruple
and $\hat\imath=0,\ldots,k{-}1$.
Each such subspace is identified with the algebra $\Hbb$
of quaternions realized in terms of $2{\times}2$ matrices with a basis
\begin{equation}\label{emu}
\bigl(e_{\mu_{\hat\imath}}\bigr) \=
\bigl(e^{}_{\mu_0}\bigr)\=
\bigl(-\ic\,\s_1\,, -\ic\,\s_2\,, -\ic\,\s_3\,,\,{\bf1}_2\bigr)\ ,
\end{equation}
where $\s_\a$, $\a{=}1,2,3$, are the Pauli matrices.

Defining $V_\mu =\de_{\mu\nu}V^\nu = (\de_{\mu\nu}V^\nu_{aP})$, one can write
$V_\mu$ with $(\mu)=(\mu_0,\ldots,\mu_{k-1})$ as a collection of $k$
quaternionic $k{\times}1$ column vectors, which we choose as follows,
\begin{equation}\label{Vmu}
V_{\mu_0} = \begin{pmatrix}
e^\+_{\mu_0}\\{\bf 0}_2\\\vdots\\\vdots\\{\bf 0}_2\end{pmatrix}
\ ,\quad\ldots\ ,\quad
V_{\mu_{\hat\imath}} = \begin{pmatrix}
{\bf 0}_2\\\vdots\\ e^\+_{\mu_{\hat\imath}}\\\vdots\\{\bf 0}_2\end{pmatrix}
\ ,\quad\ldots\ ,\quad
V_{\mu_{k-1}} = \begin{pmatrix}
{\bf 0}_2\\\vdots\\\vdots\\{\bf 0}_2\\e^\+_{\mu_{k-1}} \end{pmatrix}
\end{equation}
with $e^\+_{\mu_{\hat\imath}}$ in the $\hat\imath$-th position. With the help
of these matrices we can introduce the complex $2k{\times}2$ matrix
\begin{equation}\label{bfx}
{\bf x}\=x^\mu V_\mu\=x^{\mu_0} V_{\mu_0}+\ldots +x^{\mu_{k-1}}V_{\mu_{k-1}}\=
\begin{pmatrix}x^{\mu_0}e^\+_{\mu_0}\\\vdots\\ x^{\mu_{k-1}}e^\+_{\mu_{k-1}}
\end{pmatrix}\ =:\ \begin{pmatrix}x_0\\\vdots\\ x_{k-1}\end{pmatrix}\ .
\end{equation}

Note that (\ref{x}), (\ref{V}), (\ref{ds}) and (\ref{bfx}) are 
invariant only under the subgroup Sp(1)${}\times{}$Sp$(k)/{\Z}_2$ of the 
$\R^{4k}$ rotational isometry group SO$(4k)$.

\bigskip

\section{Generalized self-duality}

\noindent
The matrices (\ref{emu}) enjoy the properties
\begin{equation}
\label{prpts}
\begin{aligned}
e^\+_{\mu_{\hat\imath}}e^{}_{\nu_{\hat\jmath}}\= e^\+_{\mu^{}_0}e^{}_{\nu^{}_0}\=&
\de^{}_{\mu^{}_0\nu^{}_0}\ {\bf 1}_2 + \eta^{\a}_{\mu^{}_0\nu^{}_0}\ 
\ic\,\s_\a\ =:\ 
\de^{}_{\mu^{}_0\nu^{}_0}\ {\bf 1}_2 + \eta_{\mu^{}_0\nu^{}_0}\ , \\
e^{}_{\mu_{\hat\imath}}e^\+_{\nu_{\hat\jmath}}\= e^{}_{\mu^{}_0}e^\+_{\nu^{}_0}\=&
\de^{}_{\mu_0^{}\nu^{}_0}\ {\bf 1}_2 + \bar\eta^{\a}_{\mu^{}_0\nu^{}_0}\ 
\ic\,\s_\a\ =:\ 
\de^{}_{\mu^{}_0\nu^{}_0}\ {\bf 1}_2 + \bar\eta_{\mu^{}_0\nu^{}_0}\ ,
\end{aligned}
\end{equation}
where $\eta^{\a}_{{\mu^{}_0}{\nu^{}_0}}$ and $\bar\eta^{\a}_{{\mu^{}_0}{\nu^{}_0}}$ 
denote the self-dual and anti-self-dual 't~Hooft tensors, respectively, 
with $\mu^{}_0$ and $\nu^{}_0$ running from 1 to 4.
Due to the identities (\ref{prpts}) we have
\begin{equation}\label{VV}
V_\mu^\+ V_{\nu}+V_{\nu}^\+V_\mu\=2\,\de_{\mu\nu}\ {\bf 1}_2\ .
\end{equation}

Let us define antihermitian $2k{\times}2k$ matrices~\cite{CGK}
\begin{equation}\label{N}
N_{\mu\nu}\ :=\ \sfrac{1}{2}(V_\mu V_{\nu}^\+ - V_{\nu}V_\mu^\+)\ \phantom{,}
\end{equation}
and antihermitian $2{\times}2$ matrices
\begin{equation}\label{Nbar}
\bar N_{\mu\nu}\ :=\ \sfrac{1}{2}(V_\mu^\+ V_{\nu} - V_{\nu}^\+V_\mu )\ ,
\end{equation}
which in the four-dimensional case (when $k{=}1$) coincide with the matrices
$\eta_{\mu\nu}$ and $\bar\eta_{\mu\nu}$, respectively. Note that for
any $\mu ,\nu=1,\ldots,4k$ we have
$N_{\mu\nu}\in sp(k)\subset u(2k)$ and $\bar N_{\mu\nu}\in sp(1)\subset u(2)$.
These matrix-valued antisymmetric tensors are invariant under
Sp(1)${}\times{}$Sp$(k)/{\Z}_2$ (acting on spacetime indices) since this 
subgroup preserves the quaternionic structure on $\R^{4k}\cong\Hbb^k$. There
exists a third Sp(1)${}\times{}$Sp$(k)/{\Z}_2$ invariant tensor $M_{\mu\nu}$
taking values in the complement of $sp(1)\oplus sp(k)$ in $so(4k)$.
Its explicit form can be found e.g.~in~\cite{CGK, P}.

Let us further define a tensor
\begin{equation}\label{Q}
Q_{\mu\nu\r\s}\ :=\ \tr (V_\mu^\+ V_{\nu}V_{\r}^\+V_\s)
\end{equation}
and its total antisymmetrization
\begin{equation}\label{T}
T_{\m\n\r\s}\ :=\ \sfrac{1}{12}Q_{\m[\n\r\s ]}\= \sfrac{1}{12}
(Q_{\m\n\r\s} + Q_{\m\s\n\r} + Q_{\m\r\s\n}
-Q_{\m\r\n\s} - Q_{\m\n\s\r} - Q_{\m\s\r\n})\ ,
\end{equation}
which generalizes $\ve_{\mu\nu\r\s}$ from four to $4k$ dimensions.
By direct calculation one finds~\cite{CGK} that the matrices $N_{\mu\nu}$
are self-dual in this generalized sense,
\begin{equation}\label{TNN}
\sfrac{1}{2}T_{\mu\nu\r\s}N_{\r\s}\=N_{\mu\nu} \ ,
\end{equation}
while
\begin{equation}\label{TNbar}
\sfrac{1}{2}T_{\mu\nu\r\s}\bar N_{\r\s}\=- \sfrac{2k+1}{3}\bar N_{\mu\nu}\ .
\end{equation}

\bigskip

\section{Linear system and generalized self-dual Yang-Mills equations}

\noindent
We consider a gauge potential $A=A_\m\diff x^\m$ and the Yang-Mills field
$F=\diff A+A\wedge A$ with components $F_{\m\n}=\pa_\m A_\n - \pa_\n A_\m
+ [A_\m ,A_\n ]$,  where $(x^\m)\in\R^{4k}$ and $\pa_\m :=\pa /\pa x^\m$. 
The fields $A_\m$ and $F_{\m\n}$ take values in the Lie algebra $u(2r)$. 

Let us consider the linear system of equations
\begin{equation}\label{linsys}
\la^P V^\m_{aP}(\pa_\m\j + A_\m\j )\=0\qquad\mbox{for}\quad a=1,\ldots,2k\ ,
\end{equation}
where $\la^P$ with $P=1,2$ are homogeneous coordinates on the Riemann sphere
$\C P^1 =\C\cup\{\infty\}$, the matrix elements $V^\m_{aP}$ are given in 
(\ref{Vmu}), and the auxiliary $2r{\times}2r$ matrix $\psi$ depends on $x^\mu$
and $\la^P$.
The compatibility conditions of this linear system read
\begin{equation}\label{comp}
(V^\m_{aP}V^\n_{bQ}+V^\m_{aQ}V^\n_{bP} )F_{\m\n}\=0 \ .
\end{equation}
Using the definitions of $V_\m, Q_{\m\n\r\s}$ and $T_{\m\n\r\s}$, we find that
these conditions are equivalent~\cite{W,CGK} to the equations
\begin{equation}\label{gsdym}
\sfrac{1}{2}T_{\mu\nu\r\s} F_{\r\s} \= F_{\mu\n}\ .
\end{equation}
In four dimensions $T_{\mu\nu\r\s}=\ve_{\mu\nu\r\s}$ and hence (\ref{gsdym})
coincides with the standard self-dual Yang-Mills equations. In higher
dimensions these equations can be considered as generalized SDYM equations. 
Obviously, any gauge field fulfilling (\ref{gsdym}) also satisfies the 
second-order Yang-Mills equations due to the Bianchi identities.

Solutions to (\ref{gsdym}) arising from the linear system (\ref{linsys}) 
can be constructed via a (generalized) twistor approach~\cite{W}. 
Namely, one can introduce a twistor space $\R^{4k}\times\C P^1$ of $\R^{4k}$.
As a complex manifold, this space is the holomorphic vector bundle
\begin{equation}\label{hvb}
{\Pcal}^{2k+1}\ \to\ \C P^1 \qquad\textrm{with}\qquad
{\Pcal}^{2k+1}\={\cal O}(1)\otimes\C^{2k} \ .
\end{equation}
More precisely, ${\Pcal}^{2k+1}$ is an open subset of $\C P^{2k+1}$ which is 
the twistor space of $\Hbb P^k\supset \Hbb^k=\R^{4k}$. 
With the help of the linear system (\ref{linsys}), one can show~\cite{W}
that there is a one-to-one correspondence between gauge equivalence classes of
solutions to the self-duality equations (\ref{gsdym}) and equivalence classes
of holomorphic vector bundles $E$ over $\Pcal^{2k+1}$ such that their 
restriction to any curve $\C P^1_x\hookrightarrow\Pcal^{2k+1}$ is trivial. 
Employing this correspondence, one can apply Ward's splitting method
\cite{Ward,W} for obtaining solutions to the self-duality equations
(\ref{gsdym}).

Instead of (\ref{gsdym}) we might ponder about (cf.(\ref{TNbar}))
\begin{equation}\label{gasdym}
\sfrac{1}{2}T_{\mu\nu\r\s}F_{\r\s}\=- \sfrac{2k+1}{3}F_{\mu\nu}
\end{equation}
as generalized anti-self-duality equations. When $k{\ge}2$ however,
these equations cannot arise as compatibility conditions for a 
linear system~\cite{CGK}. Note that if one changes the definitions
(\ref{Vmu}) by substituting $e^\+_{\mu_{\hat\imath}}$ with 
$e_{\mu_{\hat\imath}}$ and redefines $T_{\mu\nu\r\s}\to-T_{\mu\nu\r\s}$,
then (\ref{gsdym}) and (\ref{gasdym}) (each with a sign change) will generalize
the four-dimensional anti-self-duality and self-duality equations, respectively
\cite{P}. Yet, this does not change the fact that only (\ref{gsdym}) arises 
from the integrability conditions (\ref{comp}) of the linear system 
(\ref{linsys}).

\bigskip

\section{Extended ADHM construction in 4k dimensions}

\noindent
Recall that we consider gauge fields with values in the Lie algebra 
$u(2r)\supset sp(r)$.
A simple extension of the ADHM construction to $4k$ dimension~\cite{CGK}
for generating solutions to~(\ref{gsdym}) is based on
\begin{equation} \label{Psi}
\textrm{a }\quad(2\ell{+}2r)\times 2r\quad \textrm{matrix}\quad \Psi
\qquad\textrm{and}
\end{equation}
\begin{equation}\label{Delta}
\textrm{a }\quad(2\ell{+}2r)\times2\ell \quad\textrm{matrix }\quad
\Delta\={\bf a}+{\bf b}\ ({\bf x} {\otimes} {\bf1}_\ell)\=
{\bf a} + \sum\limits^{k-1}_{\hat\imath =0}{\bf b}_{\hat\imath}\ 
(x_{\hat\imath} {\otimes} {\bf1}_\ell) \ ,
\end{equation}
where $\bf a$ and~${\bf b}_{\hat\imath}$ are constant 
$(2\ell{+}2r)\times 2\ell$ matrices.\footnote{
Note that $\bf b$ is a constant $(2\ell{+}2r)\times2k\ell$ matrix,
$\bf x$ is the $2k\times 2$ matrix given in~(\ref{bfx}) and
$x_{\hat\imath}=x^{\m_{\hat\imath}}_{}e^\+_{\m_{\hat\imath}}$ is $2\times 2$.
Correspondingly, ${\bf x}\otimes{\bf 1}_\ell$ and 
$x^{}_{\hat\imath}\otimes{\bf 1}_\ell$ are $2k\ell\times 2\ell$ and 
$2\ell\times 2\ell$ matrices, respectively.}
The matrices (\ref{Psi}) and (\ref{Delta}) are subject to the following 
conditions:
\begin{align}
\Delta^\+\Delta\quad& \textrm{is invertible}\ ,\label{c1}\\
[\,\Delta^\+\Delta\,,\,V_\mu\otimes{\bf1}_\ell\,]\ &=\ 0 \ ,\label{c2}\\
\Delta^\+\Psi\ &=\ 0\ ,\label{c3}\\
\Psi^\+\Psi\ &=\ {\bf1}_{2r} \ ,\label{c4}\\
\Psi\,\Psi^\+\ +\ \Delta\,(\Delta^\+\Delta)^{-1}\Delta^\+\ &=\ 
{\bf1}_{2\ell + 2r}\ .
\label{complete}
\end{align}
The completeness relation (\ref{complete}) means that $\Psi\,\Psi^\+$ and 
$\Delta\,(\Delta^\+\Delta)^{-1}\Delta^\+$ are projectors on orthogonal
complementing subspaces of $\C^{2\ell +2r}$.

{}For $(\Delta,\Psi)$ satisfying~(\ref{c1})--(\ref{complete})
the gauge potential is chosen in the form
\begin{equation} \label{adhmA}
A\ =\ \Psi^\+\,\diff\Psi\quad\Rightarrow\quad A_\mu\ =\ \Psi^\+\,\pa_\mu\Psi\ .
\end{equation}
The components of the gauge field $F$ then take the form
\begin{equation} 
\label{adhmF}
\begin{aligned}
F_{\mu\nu}\ &=\
\pa_\mu(\Psi^\+\pa_\nu\Psi)\ -\ \pa_\nu(\Psi^\+\pa_\mu\Psi)\
+\ [\,\Psi^\+\pa_\mu\Psi\,,\,\Psi^\+\pa_\nu\Psi\,] \\
&=\ \Psi^\+\bigl\{(\pa_\mu\Delta )(\Delta^\+ \Delta )^{-1}\pa_\nu\Delta^\+ -
(\pa_\nu\Delta )(\Delta^\+ \Delta )^{-1}\pa_\mu\Delta^\+\bigr\}\Psi \\
&=\ 2\,\Psi^\+ {\bf b}\,N_{\mu\nu}(\Delta^\+\Delta)^{-1} \,{\bf b}^\+\Psi\ .
\end{aligned}
\end{equation}
{}From  (\ref{c2}) and (\ref{adhmF}) one easily sees that 
$F_{\mu\nu}^\+=-F_{\mu\nu}$. Due to (\ref{TNN}) the gauge field (\ref{adhmF}) 
indeed solves the generalized SDYM equations (\ref{gsdym}).

\bigskip

\section{Self-dual noncommutative deformation}

\noindent
A Moyal deformation of Euclidean~$\R^{4k}$ is achieved by replacing
the ordinary pointwise product of functions on it by the nonlocal
but associative Moyal star product. The latter is characterized by
a constant antisymmetric matrix $(\th^{\mu\nu})$ which prominently
appears in the star commutation relation between the coordinates,
\begin{equation} \label{nccoord}
[\,x^\mu\,,\,x^\nu\,] \= \ic\,\th^{\mu\nu} \ .
\end{equation}
A different realization of this algebraic structure keeps the standard
product but promotes the coordinates (and thus all their functions)
to noncommuting operators acting in an auxiliary Fock space~$\Hcal =
\Hcal_0\otimes\Hcal_1\otimes\ldots\otimes\Hcal_{k-1}$, where each
$\Hcal^{}_{\hat\imath}$ is a two-oscillator Fock space~\cite{Sei}.
The two formulations are tightly connected through the Moyal-Weyl map.
When dealing with noncommutative U($2r$) Yang-Mills theory from now on,
we shall not denote the noncommutativity by either inserting stars in all
products or by putting hats on all operator-valued objects, but simply
by agreeing that our coordinates are subject to~(\ref{nccoord}).
The existence of~$(\th^{\mu\nu})$ breaks the Euclidean SO($4k$) symmetry
but we may employ SO($4k$) rotations to go to a basis in which $(\th^{\mu\nu})$
takes Darboux form, i.e.~the only nonzero entries are 
\begin{equation} 
\th^{4\hat\imath+1,4\hat\imath+2}\=-\th^{4\hat\imath+2,4\hat\imath+1}
\qquad\textrm{and}\qquad
\th^{4\hat\imath+3,4\hat\imath+4}\=-\th^{4\hat\imath+4,4\hat\imath+3}
\qquad\textrm{for}\quad\hat\imath=0,\ldots,n \ ,
\end{equation}
where we defined $n:=k{-}1$ for convenience.
Such a matrix is block-diagonal, with each $4{\times}4$ block labelled by
$\hat\imath$ being some linear combination of the self-dual 't~Hooft tensor 
$(\h^{3\mu_0\nu_0})$ and the anti-self-dual one $(\bar\h^{3\mu_0\nu_0})$.
In this letter we shall restrict ourselves to the purely self-dual situation,
\begin{equation} \label{comrel}
\th^{\m_{\hat\imath}\n_{\hat\imath}}\=\th_{\hat\imath}\,\h^{3\m_0\n_0}
\qquad\textrm{and zero otherwise}\ ,
\end{equation}
which is characterized by real numbers $(\th_0,\th_1,\ldots,\th_n)$.
In this case the extended ADHM construction of the previous section survives
the deformation unharmed.

\bigskip

\section{Noncommutative 't~Hooft-type solution in 4k dimensions}

\noindent
For making contact with the results of~\cite{ILM}, 
we choose in the entries of the noncommutativity matrix~(\ref{comrel}) as 
\begin{equation} \label{theta}
\th_0 \= -\th_i \ =:\ \th \qquad\textrm{for}\quad i=1,\ldots,n\ .
\end{equation}
Let us pick the gauge group U(2) (i.e.~$r{=}1$) and take\footnote{
The same ansatz can be considered for the gauge group U$(2r)$. 
For a more general 't~Hooft-type ansatz see~\cite{CGK}.} (see (\ref{Delta}))
\begin{equation} \label{anz}
{\bf a}\ =\ \begin{pmatrix}
\La_1{\bf1}_2 & \ldots & \La_n{\bf1}_2 \\
{\bf0}_2         &        & {\bf0}_2      \\
              & \ddots &               \\
{\bf0}_2      &        & {\bf0}_2          \end{pmatrix}
\ ,\qquad
{\bf b}^{}_{\hat\imath}\ =\ \begin{pmatrix}
{\bf0}_2 & \ldots & {\bf0}_2 \\
b^{1}_{\hat\imath}{\bf1}_2 &        & {\bf0}_2 \\
         & \ddots &          \\
{\bf0}_2 &        & b^{n}_{\hat\imath}{\bf1}_2 \end{pmatrix}
\qquad\textrm{and}\qquad
\Psi\ =\ \begin{pmatrix}
\Psi_0 \\ \Psi_1 \\ \vdots \\ \Psi_n \end{pmatrix}\ ,
\end{equation}
where $\La_i$ and $b^{i}_{\hat\jmath}$ are real constants and 
we have put $\ell =n$ (see section 5). Moreover, we choose
\begin{equation} \label{b}
b_0^i=1 \quad\textrm{for}\quad i=1,\ldots,n 
\qquad\textrm{and}\qquad b_1^1=\ldots=b^n_n=-1
\qquad\textrm{but}\qquad b^{i}_{\hat\jmath}=0 \quad\textrm{otherwise}\ .
\end{equation}
With this selection we obtain
\begin{equation} \label{deltaform}
{\D}\ =\ \begin{pmatrix}
\La_1{\bf1}_2 & \ldots & \La_n{\bf1}_2 \\
\tilde x_1           &        & {\bf0}_2      \\
              & \ddots &               \\
{\bf0}_2      &        & \tilde x_n          \end{pmatrix}
\qquad\textrm{and}\qquad
{\D^\+}\ =\ \begin{pmatrix}
\La_1{\bf1}_2 & \tilde x_1^\+   &        & {\bf0}_2 \\
   \vdots     &          & \ddots &          \\
\La_n{\bf1}_2 & {\bf0}_2 &        & \tilde x_n^\+   \end{pmatrix}\ ,
\end{equation}
where
\begin{equation}\label{tilde_x}
\tilde x_i\ :=\ x_0 - x_i \qquad\textrm{for}\quad i=1,\ldots,n\ .
\end{equation}

Using (\ref{prpts}) and (\ref{comrel})--(\ref{theta}), we find that
\begin{equation}\label{prod_x}
\tilde x_i^\+ \tilde x_i \=\tilde x_i\,\tilde x_i^\+ \=\tilde r^2_i \,{\bf1}_2
\qquad\textrm{with}\quad \tilde r^2_i\=\tilde x_i^\m \tilde x_i^\m
\qquad\qquad\textrm{(no sum over $i$)} \ .
\end{equation}
At this point we observe that (\ref{comrel})--(\ref{theta}), (\ref{prod_x}) 
and (\ref{deltaform})--(\ref{tilde_x}) coincide with formulae 
(4.1), (4.3) and (2.18) from~\cite{ILM} (where noncommutative
't~Hooft instantons in four dimensions were discussed) 
if we identify our $(x_0,x_i)$ with their $(x,a_i)$. 
Therefore, all computations in~\cite{ILM} also extend to our case. 
In particular, the operator $\tilde r^2_i$ is invertible
on the Fock space $\Hcal$, and the task to solve (\ref{c3}) and (\ref{c4}) 
is accomplished with (cf.~(4.9) of~\cite{ILM})
\begin{equation} \label{Psisol}
\Psi_0\ =\ \phi_n^{-\frac12}\,{\bf1}_2 \qquad\textrm{and}\qquad
\Psi_i\ =\ -\tilde x_i\,\frac{\La_i}{\tilde r_i^2}\,\phi_n^{-\frac12}\ ,
\end{equation}
where
\begin{equation}\label{phi_n}
\phi_n\ =\
1+\ \sum_{i=1}^n\frac{\La_i^2}{\tilde r_i^2} \ .
\end{equation}
{}Furthermore, by direct calculation one can show that the completeness
relation (\ref{complete}) is satisfied. Therefore, we can define a gauge 
potential via (\ref{adhmA}) and obtain from (\ref{adhmF}) a self-dual
gauge field on $\R^{4k}_{\th}$.

We remark that the fields $\ A_{\mu_0}=\Psi^\+\,\pa_{\mu_0}\Psi\ $ and
\begin{equation}\label{AandF}
F_{\mu_0\nu_0}\=
2\Psi^\+ {\bf b}\,N_{\mu_0\nu_0}(\Delta^\+\Delta)^{-1}{\bf b}^\+\Psi\=
2\Psi^\+ {\bf b}\,\begin{pmatrix}
\h_{\mu_0\nu_0} & \ldots & {\bf0}_2 \\
   \vdots       & \ddots & \vdots   \\
  {\bf0}_2      & \ldots & {\bf0}_2 \end{pmatrix}
(\Delta^\+\Delta)^{-1} \,{\bf b}^\+\Psi
\end{equation}
do not contain any derivatives with respect to $x_1,\ldots ,x_n$. Hence, 
(\ref{AandF}) exactly reproduces the noncommutative $n$-instanton solution 
in four dimension as derived in~\cite{ILM}, if only their translational moduli
$a_i$ are identified with our coordinates $x_i$ on $\R^{4n}_{\th}$,
which complements $\R^{4}_{\th}$ in $\R^{4k}_{\th}$ (recall $k=n{+}1$).
Since $x_0$ and $x_i$ with $i=1,\ldots,n$ are operators acting on the Fock 
space $\Hcal_0\otimes\Hcal_1\otimes\ldots\otimes\Hcal_n\otimes\C^2$ one may 
interpret the configuration (\ref{AandF}) as a quantum 't~Hooft $n$-instanton. 
Note, however, that the scale parameters $\Lambda^2_i$ entering $\Psi$ and 
(\ref{AandF}) are not quantized.

\bigskip

\section{Noncommutative BPST-type solution in eight dimensions}

\noindent
In the commutative case, as it was explained in~\cite{CGK}, the 't~Hooft-type
ansatz of the previous section produces solutions of the YM equations on 
$\R^{4k}$ which cannot be extended to the quaternionic projective space 
$\Hbb P^k$ if $k\ge 2$. Hence, these solutions cannot have finite topological 
charges beyond $k{=}1$. However, their four-dimensional ``slice" describes
't~Hooft instantons both in the commutative and noncommutative cases. We shall
now consider a different kind of ansatz for $\bf a$ and $\bf b$ in 
(\ref{Delta}) which generates true instanton-type configurations 
(with finite Pontryagin numbers) since in the commutative limit 
$\th_{\hat\imath}\to 0$ they can be extended to $\Hbb P^k$.

For simplicity, we restrict ourselves to eight dimensions ($k{=}2$) and to the 
gauge group U(4). For the commutative case this kind of ansatz was discussed
in Appendix B of~\cite{CGK}, but the matrix $\Psi$ proposed there fails
to obey (\ref{c3}). By properly solving the extended ADHM equations 
(\ref{c1})--(\ref{complete}) for the eight-dimensional U(4) setup we shall 
derive a smooth noncommutative instanton configuration which in the commutative
limit generalizes the finite action solutions on $\R^{4}$ to $\R^8$.

We consider the noncommutative space $\R^8_\th$ with coordinates
$(x^\m )=(x^{\m_0}_{}, y^{\n_0}_{})$ such that
\begin{equation}\label{x&y}
[x^{\m_0}_{}, x^{\n_0}_{}]= \ic\,\th_0\,\h^{3\m_0\n_0}\ ,\qquad
[x^{\m_0}_{}, y^{\n_0}_{}]=0\qquad\textrm{and}\qquad
[y^{\m_0}_{}, y^{\n_0}_{}]= \ic\,\th_1\,\h^{3\m_0\n_0}\ ,
\end{equation}
with $\th_0, \th_1 >0$. We introduce
\begin{equation}\label{xu}
x\=x^{\m_0}_{}e^\+_{\m_0}\ ,\qquad y\=y^{\m_0}_{}e^\+_{\m_0}
\qquad\textrm{and}\qquad
{\bf x} \= \begin{pmatrix}x\\y\end{pmatrix}\ ,
\end{equation}
which obey
\begin{equation}\label{x+y+}
x^\+ x \= x^{\m_0}_{}x^{\m_0}_{}\,{\bf1}_2 = r^2_0 \,{\bf1}_2
\qquad\textrm{and}\qquad
y^\+ y \= y^{\m_0}_{}y^{\m_0}_{}\,{\bf1}_2 = r^2_1 \,{\bf1}_2\ ,
\end{equation}
\begin{equation}\label{xx+yy+}
xx^\+\=\begin{pmatrix}r_0^2-2\th_0&0\\0&r_0^2+2\th_0\end{pmatrix}
\qquad\textrm{and}\qquad
yy^\+\=\begin{pmatrix}r_1^2-2\th_1&0\\0&r_1^2+2\th_1\end{pmatrix}\ .
\end{equation}

Let us choose
\begin{equation}\label{abbb}
{\bf a} \=\begin{pmatrix}\La{\bf1}_2\\{\bf0}_2\\{\bf0}_2\end{pmatrix}\ ,\qquad
{\bf b}_1 \=\begin{pmatrix}{\bf0}_2\\{\bf1}_2\\{\bf0}_2\end{pmatrix}\ ,\qquad
{\bf b}_2 \=\begin{pmatrix}{\bf0}_2\\{\bf0}_2\\{\bf1}_2\end{pmatrix}
\qquad\Rightarrow\qquad
{\bf b} \=\begin{pmatrix}{\bf0}_2&{\bf0}_2\\{\bf1}_2&{\bf0}_2\\
                         {\bf0}_2&{\bf1}_2\end{pmatrix}
\end{equation}
and parametrize
\begin{equation}\label{psiphi}
\Psi \= \begin{pmatrix}{\psi}_0&{\phi}_0\\{\psi}_1&{\phi}_1\\
{\psi}_2&{\phi}_2\end{pmatrix}\ ,
\end{equation}
where $\La$ is a scale parameter and all blocks ${\psi}_0,\ldots,{\phi}_2$
of the $6{\times}4$ matrix $\Psi$ are $2{\times}2$ matrices. 
Then we obtain (cf.~\cite{CGK})
\begin{equation}\label{DD+}
\Delta \= {\bf a} +{\bf b\, x} \=
\begin{pmatrix}\La{\bf1}_2\\x\\y\end{pmatrix} \quad\Rightarrow\quad
\Delta^\+ \= (\La{\bf1}_2\ x^\+\ y^\+) \quad\Rightarrow\quad
\Delta^\+\Delta \=(r^2+\La^2)\cdot{\bf1}_2\ ,
\end{equation}
where $r^2=r_0^2+r_1^2$,
and the extended ADHM equations (\ref{c3}) become
\begin{equation} \label{exadhm}
\La\psi_0 + x^\+\psi_1 + y^\+\psi_2 \=0 \qquad\textrm{and}\qquad
\La\phi_0 + x^\+\phi_1 + y^\+\phi_2 \=0\ .
\end{equation}
We choose a solution to these equations in the form
\begin{equation}\label{solPsi}
\Psi = \begin{pmatrix}x^\+&y^\+\\-\La{\bf1}_2&{\bf0}_2\\
{\bf0}_2& -\La{\bf1}_2 \end{pmatrix} \begin{pmatrix}a&b\\0&c\end{pmatrix}
=\begin{pmatrix}x^\+a&x^\+b + y^\+c\\-\La a&-\La b\\
{\bf0}_2& -\La c \end{pmatrix}\ \Rightarrow\
\Psi^\+ = \begin{pmatrix} a^\+x & -\La a^\+&{\bf0}_2\\
b^\+x+c^\+y&-\La b^\+&-\La c^\+\end{pmatrix},
\end{equation}
where the $2{\times}2$ matrices $a,b$ and $c$ are fixed by the 
normalization condition (\ref{c4}) to be
\begin{equation}\label{solabc}
\begin{aligned}
a \=& \begin{pmatrix} (r_0^2+\La^2-2\th_0)^{-\sfrac{1}{2}}_{}
&0\\0&(r_0^2+\La^2+2\th_0)^{-\sfrac{1}{2}}_{}\end{pmatrix}\ ,\qquad
b\=-a^2xy^\+c\ ,\\
c\=& \frac{(r_0^2+\La^2)^{\sfrac{1}{2}}_{}}{\La}
\begin{pmatrix} (r^2+\La^2-2\th_1)^{-\sfrac{1}{2}}_{}
&0\\0&(r^2+\La^2+2\th_1)^{-\sfrac{1}{2}}_{}\end{pmatrix}\ .
\end{aligned}
\end{equation}

To be sure that our solution (\ref{solPsi})--(\ref{solabc}) contains all zero
modes of the operator $\Delta^\+$, we should check the completeness relation
(\ref{complete}). It is known that in the noncommutative case the latter
may be violated~\cite{LP} unless an additional effort is made.
After a lengthy calculation we obtain
\begin{equation}\label{PsiPsi}
\Psi\,\Psi^\+ \= \begin{pmatrix}
\frac{r^2}{r^2+\La^2}\,{\bf1}_2&-\frac{\La}{r^2+\La^2}\,x^\+&
-\frac{\La}{r^2+\La^2}\,y^\+\\
-x\,\frac{\La}{r^2+\La^2}&{\bf1}_2-x\,\frac{1}{r^2+\La^2}\,x^\+&
-x\,\frac{1}{r^2+\La^2}\,y^\+\\
-y\,\frac{\La}{r^2+\La^2}&-y\,\frac{1}{r^2+\La^2}\,x^\+&
{\bf1}_2-y\,\frac{1}{r^2+\La^2}\,y^\+\end{pmatrix}
\end{equation}
and
\begin{equation}\label{DDDD}
\Delta\,(\Delta^\+\Delta)^{-1}\Delta^\+ \= \begin{pmatrix}
\frac{\La^2}{r^2+\La^2}\,{\bf1}_2&\frac{\La}{r^2+\La^2}\,x^\+_{}&
\frac{\La}{r^2+\La^2}\,y^\+_{}\\
x\,\frac{\La}{r^2+\La^2}&x\,\frac{1}{r^2+\La^2}\,x^\+_{}&
x\,\frac{1}{r^2+\La^2}\,y^\+_{}\\
y\,\frac{\La}{r^2+\La^2}&y\,\frac{1}{r^2+\La^2}\,x^\+_{}&
y\,\frac{1}{r^2+\La^2}\,y^\+_{}\end{pmatrix} \ ,
\end{equation}
and the latter two matrices add up to ${\bf1}_6$, as the completeness relation
(\ref{complete}) demands. So, for our solution $(\Delta,\Psi )$ of 
(\ref{c2})--(\ref{complete}) the gauge field (\ref{adhmF}) satisfies the 
self-duality equations (\ref{gsdym}) (and the full YM equations) on $\R^8_\th$.

In the commutative limit $\th_0, \th_1\to 0$ our $(\Delta ,\Psi )$ produces a 
novel solution to the generalized SDYM equations (\ref{gsdym}) on 
$\R^8\cong\Hbb^2$. In~\cite{CGK} it was argued that such kind of solutions
can be extended to the compact space $\Hbb P^2$. Briefly, the arguments are as
follows~\cite{CGK}. Homogeneous coordinates on $\Hbb P^2$ are
\begin{equation}\label{homcor}
\begin{pmatrix}z'\\x'\\y'\end{pmatrix}\in \Hbb^3\ .
\end{equation}
The patch $U_1$ on $\Hbb P^2$ is defined by the condition $\det (z')\ne 0$.
Coordinates $(x,y)=(x_1,y_1)$ on this patch are obtained from (\ref{homcor})
by multiplying with $(z')^{-1}$,
\begin{equation}\label{patch1}
U_1:\
\begin{pmatrix}z'\\x'\\y'\end{pmatrix}\ \mapsto\
\begin{pmatrix}1\\x'(z')^{-1}\\y'(z')^{-1}\end{pmatrix}=:
\begin{pmatrix}1\\x_1\\y_1\end{pmatrix} \ ,
\end{equation}
and a scale parameter $\La$ can be introduced by redefining $x_1\mapsto\La^{-1}x_1,\
y_1\mapsto\La^{-1}y_1$ and multiplying the last column in (\ref{patch1}) by $\La$.
This is the patch on which we worked. One can also consider two other patches,
which are defined by conditions $\det (x')\ne 0$ and $\det (y')\ne 0$:
\begin{equation}\label{patch23}
U_2:\
\begin{pmatrix}z'\\x'\\y'\end{pmatrix}\ \mapsto\
\begin{pmatrix}z'(x')^{-1}\\1\\y'(x')^{-1}\end{pmatrix}=:
\begin{pmatrix}x_2\\1\\y_2\end{pmatrix}
\quad\mbox{and}\quad
U_3:\
\begin{pmatrix}z'\\x'\\y'\end{pmatrix}\ \mapsto\
\begin{pmatrix}z'(y')^{-1}\\x'(y')^{-1}\\1\end{pmatrix}=:
\begin{pmatrix}x_3\\y_3\\1\end{pmatrix}
\end{equation}
with coordinates $(x_2,y_2)$ and $(x_3,y_3)$ and construct solutions
$(\Delta ,\Psi )$ and $(A,F)$ on them by reproducing all steps we performed
on the patch $U_1$. On the intersections of these patches (local) solutions 
will be glued by transition functions. For more details see~\cite{CGK}.

\bigskip

\noindent
{\bf Acknowledgements}

\medskip

\noindent
The authors are grateful to A.D.~Popov for fruitful discussions and
useful comments.
T.A.I.~acknowledges the Heisenberg-Landau Program for partial support and
the Institut f\"ur Theoretische Physik der Universit\"at Hannover for
its hospitality. This work was partially supported by the Deutsche
Forschungsgemeinschaft (DFG).

\bigskip

\end{document}